\documentclass[10pt,journal,compsoc]{IEEEtran}
\ifCLASSOPTIONcompsoc
  \usepackage[nocompress]{cite}
\else
  \usepackage{cite}
\fi
\ifCLASSINFOpdf
\else
\fi
\usepackage{amsmath,amssymb}
\usepackage{changepage}
\usepackage[flushleft]{threeparttable}
\usepackage{array,booktabs,makecell}
\usepackage{setspace}
\usepackage{algorithm}% http://ctan.org/pkg/algorithms
\usepackage{algpseudocode}% http://ctan.org/pkg/algorithmicx
\usepackage{graphicx}

\PassOptionsToPackage{hyphens}{url}\usepackage{hyperref}
\usepackage{bm}

\usepackage{accents}

\usepackage{bbm}
\usepackage{physics}
\newcolumntype{C}[1]{>{\centering\arraybackslash}p{#1}}
\newcolumntype{L}[1]{>{\raggedright\arraybackslash}p{#1}}

\usepackage{amsthm}
\usepackage{thmtools}
\declaretheoremstyle[
spaceabove=6pt, spacebelow=6pt,
headfont=\normalfont\bfseries,
notefont=\mdseries, notebraces={(}{)},
bodyfont=\normalfont,
postheadspace=0.6em,
headpunct=:
]{mystyle}

\usepackage{caption}
\usepackage{subcaption}

\usepackage{hyperref}
\usepackage{url}
\usepackage{graphicx}
\usepackage{xcolor}

\newcommand{\RN}[1]{%
  \textup{\uppercase\expandafter{\romannumeral#1}}%
}

\usepackage{array}
\usepackage{multirow}
\usepackage{amsfonts}
\usepackage{enumerate}
\usepackage{amsthm}
\usepackage{svg}
\makeatletter
\algnewcommand{\LineComment}[1]{\Statex \hskip\ALG@thistlm \(\triangleright\) #1}
\makeatother

\begin{document}
\title{Hawkes Process Modeling of Block Arrivals in Bitcoin Blockchain}

\author{Rui~Luo, Vikram~Krishnamurthy,~\IEEEmembership{Fellow,~IEEE},
        and~Erik~Blasch,~\IEEEmembership{Fellow,~IEEE}% <-this % stops a space
\IEEEcompsocitemizethanks{\IEEEcompsocthanksitem R. Luo is with the Sibley School of Mechanical and Aerospace Engineering, Cornell University, Ithaca, NY, 14850.\protect\\
E-mail: rl828@cornell.edu
\IEEEcompsocthanksitem V. Krishnamurthy is with the School of Electrical and Computer Engineering, Cornell University, Ithaca, NY, 14850.\protect\\
E-mail: vikramk@cornell.edu 
\IEEEcompsocthanksitem E. Blasch is with Air Force Office of Scientific Research (AFOSR), Arlington, VA, 22203. \protect\\
E-mail: erik.blasch.1@us.af.mil
 \IEEEcompsocthanksitem This research was supported by the  U. S. Army Research Office under grants W911NF-19-1-0365, U.S. Air Force Office of Scientific Research under grant FA9550-22-1-0016, and the National Science Foundation under grant CCF-2112457. }}

\IEEEtitleabstractindextext{
\begin{abstract}
The paper constructs a multi-variate Hawkes process model of Bitcoin block arrivals and price jumps. Hawkes processes are self-exciting point processes that can capture the self- and cross-excitation effects of block mining and Bitcoin price volatility. We use publicly available blockchain datasets to estimate the model parameters via maximum likelihood estimation. The results show that Bitcoin price volatility boost block mining rate and Bitcoin investment return demonstrates mean reversion. Quantile-Quantile plots show that the proposed Hawkes process model is a better fit to the blockchain datasets than a Poisson process model.

\end{abstract}

}

\maketitle

\IEEEdisplaynontitleabstractindextext

% \fi
\IEEEpeerreviewmaketitle

\section{Introduction}
\label{sec:introduction}
Cryptocurrencies have lately seen a surge in popularity. The most well-known one, Bitcoin, has a market capitalisation of over US$\$$780 billion. Bitcoin is more than just a virtual money as the blockchain technology that underpins it creates a transparent and shared transaction database that is a security improvement on the existing monetary exchange system. Concretely, blockchain relies on a \textit{proof-of-work} mechanism to achieve consensus and immutability. This \textit{proof-of-work} is referred to as bitcoin block mining\cite{bowden2020modeling}. The modeling of block mining, which characterizes the interaction effects between block mining and the bitcoin market, is the subject of this research.

\vspace{0.1in}
\noindent
{\bf What is blockchain mining? }

\noindent
Blockchain is a distributed digital ledger that records transactions. Each transaction is uniquely recognized by a double SHA-256 hash, and unconfirmed transactions are stored in the \textit{mempool}\footnote{Note that there is no global transaction mempool; every node keeps its own set of unconfirmed transactions.}. To secure transactions and commit them to the public ledger, blockchain uses a computational procedure known as \textit{mining}, which requires solving a cryptographic challenge. In practice, miners collect transactions from the mempool and utilize their hashes, the hash that is currently at the top of the blockchain, and a \textit{nonce} (an integer value that the miner picks at will) as inputs to the cryptographic problem. A miner is deemed to have \textit{mined} a block if it solves the problem. The miner will be rewarded (currently 6.25 bitcoins), and the block will be added to the blockchain.

\vspace{0.1in}
\noindent
{\bf Is a homogeneous Poisson process adequate? }

\noindent
In the bitcoin white paper \cite{nakamoto2008peer}, Nakamoto implicitly assumes that the number of blocks mined by a miner (attacker in the original paper) in an interval is Poisson distributed. Rosenfeld \cite{rosenfeld2014analysis} and Lewenberg et al. \cite{lewenberg2015bitcoin} used a Poisson process model to describe the block arrivals. In their analysis of the Bitcoin network, Decker and Wattenhofer \cite{decker2013information} assumed a homogeneous Poisson process, where the difficulty is constant. Kawase and Kasahara \cite{kawase2017transaction} assumed that bitcoin transactions arrive at the system according to a homogeneous Poisson process (HPP), based on which a transaction queueing model is constructed and the transaction confirmation time is examined. Despite some tail deviation, Gebraselase et al. \cite{gebraselaseJiang2021transaction} suggest that bitcoin's inter-block generation time can be effectively fitted with a negative exponential distribution, which supports the Poisson arrival assumption. According to Cao et al. \cite{cao2020performanceDAG}, the time it takes a miner $i$ ($i=1,\cdots, n$) to mine a block is exponentially distributed with mean ${\frac{r_i}{D}}$, where $r_i$ is miner $i$'s hash rate and $D$ is the target difficulty. Since the minimum of $n$ independent exponential random variables is itself an exponential random variable, the time for the first block to be mined by all miners is exponentially distributed with mean $\frac{\sum_{i=1}^{n} r_i}{D}$. As a result, the block arrivals follow a Poisson counting process. 

There is strong motivation to build on the above studies to take into account mining difficulty, and adjust the assumption that block arrives as a homogeneous Poisson process. For example,
% However, none of the aforementioned studies took into account the mining difficulty adjustment, and so the block arrival process homogeneity assumption is not robust. 
Cao et al.\cite{cao2020performanceDAG} proposed a block mining success probability model which is linked to a set of fixed parameters including $n$ (the number of miners), $\sum_{i=1}^{n} r_i$ (total hash rate), and $D$ (mining difficulty), and there is considerable motivation to extend it to a real-world scenario in which these values are subject to change. Indeed, Bowden et al.\cite{bowden2020modeling} demonstrated that the bitcoin network's total hash rate is a primary driver of the block arrival, with many factors: (1) miners switching machines; (2) bitcoin price volatility and electricity price fluctuations; (3) halving of the mining reward.
Based on blockchain data and stochastic analysis, they revealed that the homogeneous Poisson process may not adequately fit the data. They proposed a more general model for block arrivals that took hash rate and difficulty time variance into consideration. Specifically, they provided three alternatives for the HPP model: (1) parametric or empirical modeling of the hash rate function; (2) deterministic or random modeling of the difficulty changes; and (3) the absence or presence of block propagation delay. To summarize, the aim of this paper is to build on previous important works \cite{laub2022book, bacry2015hawkesFinance, bowden2020modeling} and develop a more compact model that takes into consideration all relevant elements.

\vspace{0.1in}
\noindent
{\bf Why Hawkes processes?}

\noindent
This paper uses Hawkes processes\cite{hawkes1971spectra} to model self- and cross-excitation of block mining and bitcoin market volatility. The Hawkes process allows for the triggering effects of previous events on future occurrences. The multivariate Hawkes process accounts for the mutual excitements (i.e., temporal shifts) of multiple point processes with homogeneous (constant) and inhomogeneous (shift) Poisson arrival rates. Bowsher \cite{bowsher2007modelling} utilized the Hawkes process to model the mutual excitements of the trade occurrences and the intensity of mid-price changes. Filimonov and Sornette\cite{filimonov2012reflexivity} model the mid-price changes over time as a Hawkes process. They took the estimated branching ratio of the Hawkes process to quantify the economic reflexivity. Phillips and Gorse\cite{phillips2018CryptoAndSocialMedia} employed a Hawkes process to model the relationship between cryptocurrency price changes and related topic discussion in social media. Other examples of Hawkes process applications include topic clustering\cite{du2015dirichlet}, disease network\cite{choi2015constructing}, and malicious activity detection\cite{zheng2021using}.

We make the simple but sound assumption that the majority of block arrival variations are driven by either \textit{self-exciting effects} (how block mining affects itself) or \textit{cross-exciting effects} (how block mining impacts other block miners and how the cryptocurrency market affects block mining). Hawkes processes offer an intuitive notion of exogenous and endogenous components contributing to event clustering. The exogenous component in block mining corresponds to the mining difficulty predetermined by the blockchain protocol\footnote{Mining difficulty is adjusted every 2016 blocks, or roughly every two weeks.}, while the endogenous component, referred to as "market reflexivity"\cite{filimonov2012reflexivity, hardiman2013reflexivity}, corresponds to the internal feedback processes. Internal feedback includes a variety of components, such as how miners alter their mining power depending on their assessment of overall mining activity on the blockchain (e.g., total hash rate and mining hardware market), or how miners respond to cryptocurrency price fluctuations and reward halving.
% i.e., how block miners adjust their mining power based on the mining activity in the block and cryptocurrency price fluctuation. 

In this paper, we consider the following questions by using the Hawkes process as a modeling tool for the block arrival process:
\begin{list}{\labelitemi}{\leftmargin=0em}
\item \textit{Does a Hawkes process fit the block arrivals better than a homogeneous Poisson process? }
\item \textit{What insights may the Hawkes process fit result bring, such as whether bitcoin price fluctuations influence block mining rates?}
\end{list}

\vspace{0.05in}
\noindent
{\bf Main Results and Organization: }

\noindent(1) Section \ref{sec:model} discusses the multivariate Hawkes process model. We focus on Hawkes processes with exponential kernel functions and show how maximum likelihood estimation can be used to efficiently determine the parameters. 

\noindent(2) Section \ref{sec:blockchain} describes the dataset utilized and preprocessed for the blockchain analysis. We construct a trivariate Hawkes process where each process corresponds to block arrival, positive price jump, or negative price jump\footnote{The idea of treating both directions of extreme price changes as events of different types in Hawkes process model has been explored in e.g. \cite{liniger2009multivariateThesis} (Section 1.5).} respectively. We fit the trivariate Hawkes process to the bitcoin dataset. The model's kernel function illustrates the exciting effects of block mining and bitcoin price volatility. We used the goodness of fit test to show that the proposed Hawkes process model fit the block arrival process better than a Poisson process model. 

\section{Background} 
\label{sec:model}
In this section, we introduce several important concepts of the multivariate Hawkes processes, including the conditional intensity function, kernel function, and compensator. We go through the different kinds of kernel functions and show how an exponential kernel function can assist with efficient likelihood computation. The Hawkes process kernel function, in particular, is utilized to connect the cross-exciting effects of block mining and the crypto market.

\subsection{Multivariate Hawkes Process and Conditional Intensity} \label{subsec: CIF}
This subsection describes the conditional intensity function, which can be used to characterise a point process. 

Point processes are probabilistic models of events in a mathematical space that are commonly used to represent event occurrences over time or space (or both)\cite{verma2021selfmulticellular}.
The conditional intensity function $\lambda(\cdot)$, which expresses the expected infinitesimal rate at which events occur, can be used to define a point process\cite{schoenberg2010introduction}.

Consider a collection of $m$ point processes $\mathbf{N}(\cdot) = (N_1(\cdot), \cdots, N_m(\cdot))$ with a joint conditional intensity $\bm{\lambda}(\cdot) = (\lambda_1(\cdot), \cdots, \lambda_m(\cdot))$. 
We denote $H(t)=\{(t_k, d_k) \}_{k=1}^{n(t)}$ an event sequence of time $t_k$ associated with event type $d_k$, where $n(t)=\sum_{i=1}^{m} N_i(t)$ denotes the total number of events of all types. Then the conditional relationship is
\begin{equation}
	P(N_i(t+h)-N_i(t)=m | H(t)) =
	\begin{cases}
		1 - \lambda_i(t) h + o(h), & m=0 \\
		\lambda_i(t)h + o(h), & m=1 \\
		o(h), & m>1 
	\end{cases}
\end{equation}
where $\lim\limits_{h\rightarrow 0} \frac{o(h)}{h} = 0$ (i.e., $o(h)$ is "little-o" of $h$).

The interactions of the $m$ point processes can be modeled by a multivariate Hawkes process. 
For component $N_i(\cdot)$, its conditional intensity has the form
\begin{equation} \label{eq: multi basic}
    \lambda_i(t) = \mu_i + \sum_{j=1}^{m} \int_{0}^{t} \phi_{ij}(t-s) dN_j(s), \quad i=1,\cdots,m
\end{equation}
where $\mu_i > 0$ is the background rate and $\phi_{ij}(\cdot)$ is the kernel function representing the effects of process $N_j$ has on process $N_i$. The integral in (\ref{eq: multi basic}) is interpreted as a Stieltjes integral \cite{dresher2012mathematicsStieltjesIntegral}.

A popular choice for $\phi_{ij}(\cdot)$ is an exponential function $\phi_{ij}(t) = \alpha_{ij} e^{-\beta_{ij} t}$, and Eq. (\ref{eq: multi basic}) can be rewritten as
\begin{equation} \label{eq: multi Exp 1}
    \lambda_{i}(t) = \mu_i + \sum_{j=1}^{m} \int_{0}^{t} \alpha_{ij} e^{-\beta_{ij} (t-s)} dN_j(s), \quad i=1,\cdots, m
\end{equation}

Another useful choice for $\phi_{ij}(\cdot)$ is a power law function $\phi_{ij}(t) = \alpha_{ij}(c_{ij} + t)^{-\beta_{ij}} $ with $c_{ij}>0, \beta_{ij}>1$, and
\begin{equation}
    \lambda_{i}(t) = \mu_i + \sum_{j=1}^{m} \int_{0}^{t} \alpha_{ij} (c_{ij} + (t-s))^{-\beta_{ij}} dN_j(s), \quad i=1,\cdots, m
\end{equation}
Compared to the exponential function, the power law kernel functions capture long term memory which is prominent in data in financial markets\cite{mark2020quantifyingbitcoin}.

\subsection{Compensator}
The integral of the conditional intensity function over time is known as the compensator of the point process, defined as
\begin{equation} \label{eq: compensator basic}
    \Lambda_i(t) = \int_{0}^{t} \lambda_i(s) ds, \quad i=1,\cdots,m
\end{equation}
Lemma 7.2.$\RN{5}$ from Daley and Vere-Jones\cite{daley2003introduction} states that the difference between the point process and the compensator, i.e., $N_i(t)-\Lambda_i(t)$, is a martingale.
For point processes with exponential kernel function (\ref{eq: multi Exp 1}), the corresponding compensator is
\begin{equation} \label{eq: compensator Exp}
\begin{split}
    \Lambda_i(t) 
    & = \mu_i t + \sum_{t_k < t} \frac{1}{\beta_{id_{k}}} \alpha_{id_{k}} (1 - e^{-\beta_{id_{k}}(t-t_k)}), \quad i=1,\cdots, m
\end{split}
\end{equation}

The compensator (\ref{eq: compensator Exp}) is useful in parameter estimation and goodness of fit testing\cite{laub2022book}. 

\subsection{Maximum Likelihood Estimation} \label{subsec: mle}
Based on Proposition 7.2.$\RN{3}$ from Daley and Vere-Jones\cite{daley2003introduction}, for point processes $\bm{N}(t)= (N_1(t), \cdots, N_m(t))$ with a joint conditional intensity $\bm{\lambda}(t) = (\lambda_1(t), \cdots, \lambda_m(t)) $ and a collection of compensators $\bm{\Lambda}(t) = (\Lambda_1(t), \cdots, \Lambda_m(t))$, 
% and the observed arrival times $\{t_1, \cdots, t_{n(T)} \}$ over the time period $[0, T]$, 
then the likelihood function $L$ for the point process is
\begin{equation} 
    L = \Big[\prod_{k=1}^{n(T)} \lambda_{d_k} (t_k) \Big] e^{-\sum_{i=1}^{m} \Lambda_i(T)}
\end{equation}
The log-likelihood $l$ is
\begin{equation} \label{eq: multi log-likelihood}
    l = \sum_{k=1}^{n(T)} \textrm{ln}(\lambda_{d_k} (t_k)) - \sum_{i=1}^{m} \Lambda_i (T)
\end{equation}

The two-tuple $(\mathbf{N}(t), \bm{\lambda}(t))$ is a Markov process if the exponential kernel's decay parameter is constant (Proposition 2 in \cite{bacry2015hawkesFinance}), i.e., $\beta_{ij}=\beta$ in (\ref{eq: multi Exp 1}). The Markov property also leads to an efficient computation of the log-likelihood (\ref{eq: multi log-likelihood}) \cite{bompaire2019machineThesis}. The log-likelihood requires evaluating $\lambda_{d_k}(t_k)$ for $k=1,\cdots, n(T)$. By utilizing the following relationship based on the Markov property,
\begin{equation}
    \bm{\lambda}(t) = \bm{\lambda} + (\bm{\lambda}(t_k) + \bm{\alpha}_{d_k} - \bm{\lambda}) 
    e^{- \beta (t-t_k)}
\end{equation}
where $\bm{\lambda} = (\lambda_1, \cdots, \lambda_m)^T, \bm{\alpha}_i = (\alpha_{i1}, \cdots, \alpha_{im})^T$, we can evaluate these intensities in $\mathcal{O}(n(T))$ instead of $\mathcal{O}(n(T)^2)$. Numerical optimization such as the Newton-style methods \cite{laub2022book} and EM algorithm \cite{veen2008estimationEM} can be used to compute the maximum likelihood estimation.

One limitation of exponential kernels is that the quality of the modeling depends on the choice of $\beta$. The sum of exponentials kernels, defined as $\phi_{ij}(\cdot) = \sum_{u=1}^{U} a_{ij}^{u} e^{-\beta^{u}t}$, include several decay parameters $\beta^u$ of various time scales, making the modeling less susceptible to initial choice of $\beta$. Sum of exponentials kernels can also be used to approximate power-law kernels with suitably chosen parameters $\beta^u,u=1,\cdots,U$\cite{bompaire2019machineThesis}. The corresponding conditional intensity functions are
\begin{equation} \label{eq: intensity sumexp 1}
    \lambda_i(t) = \mu_i + \sum_{u=1}^{U} v_i^u(t), \quad i=1,\cdots,m
\end{equation}
where
\begin{equation} \label{eq: intensity sumexp 2}
    v_i^u(t) = \sum_{j=1}^{m} \int_0^t \alpha_{ij}^u e^{-\beta^u (t-s)} dN_j(s)
\end{equation}
The corresponding compensators (\ref{eq: compensator basic}) are 
\begin{equation} \label{eq:compensator multi sum}
    \Lambda_i(t) = \mu_i t + \sum_{j=1}^{m} \sum_{u=1}^{U} \alpha_{ij}^{u} M_j^u(t), \quad i=1,\cdots,m
\end{equation}
where 
\begin{equation}
\begin{split}
    % M_j^u(t) & = \int_0^t e^{-\beta^u (s-t_k)} dN_j(s) \\ 
    M_j^u(t) &= \int_{0}^{t} \sum_{\substack{ t_k<s, \\ d_k=j}} e^{-\beta^u (s-t_k)} ds \\
    &= \sum_{\substack{t_k < t,\\d_k=j}} \frac{1}{\beta^u} (1-e^{-\beta^u (t - t_k)})
\end{split}
\end{equation}

In Section \ref{subsec: block and crypto}, we will use Hawkes processes with the sum of exponentials kernel to model the self- and cross-exciting effects between block mining and crypto market.

\section{Blockchain Data Analysis}
\label{sec:blockchain}
In this section, we analyze blockchain datasets by constructing a multivariate Hawkes process model for block arrivals and bitcoin price jumps.  We estimate the parameters based on maximum likelihood estimation technique summarized in Section \ref{subsec: mle}. The derived kernel functions are utilized as a quantification of how block mining and the bitcoin market interact.

\subsection{Datasets} \label{subsec: data}
Our data is comprised of two parts: blockchain data and bitcoin price data. Blockchain data is obtained from Google Bigquery \cite{bigquery} and includes timestamps of all the blocks mined between January 1 and March 1 of 2022. We discuss how we deal with erroneous block timestamps in Appendix \ref{append:cleaning}.
Bitcoin price data is obtained using an open source API\cite{crypto-chassis}, which includes the processed volume-weighted average price (VWAP)\footnote{We use VWAP instead of open/high/low/close prices for its better measurement of the bitcoin market value.} for 5-minute intervals since January 1 of 2016 in Coinbase, one of the largest crypto exchanges.

To construct Hawkes processes corresponding to bitcoin price changes, we need to obtain "events" data from the bitcoin price time series. We use a similar approach to \cite{liniger2009multivariateThesis} and look at extreme values only, which are most likely caused by relevant external events instead of market noise.
Concretely, we first convert the price data to log returns, similar to \cite{yang2018applicationsLogReturn, embrechts2011multivariateLogReturn, phillips2018CryptoAndSocialMedia}. We then devise an upper threshold and a lower threshold. We only keep values that are above or below these two thresholds from the price time series, which correspond to positive and negative price jump events respectively. As opposed to the time series data, the time intervals between the events are not fixed in length. \cite{liniger2009multivariateThesis} uses fixed thresholds, which are the empirical $10\%$ and $90\%$ quantiles of the price time series data. 
\begin{figure}
     \centering
     \begin{subfigure}[b]{0.5\textwidth}
        \centering
        \includegraphics[width=1\textwidth]{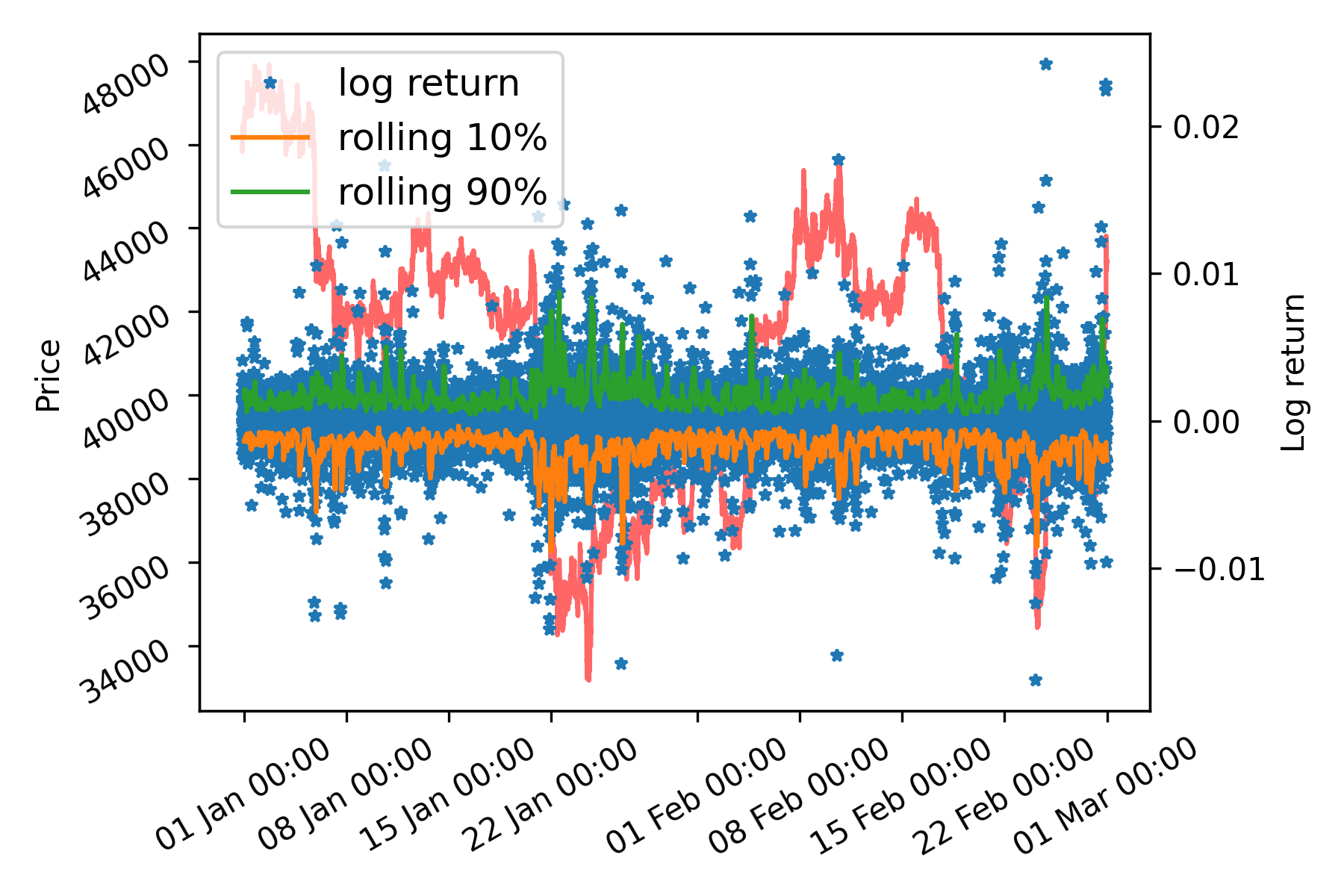}
        \caption{The bitcoin volume-weighted average price (vwap) time series (red curve), the log return (blue stars), and the log return rolling $10\%/ 90\%$ quantiles (orange curve and green curve). As mentioned in Section \ref{subsec: data}, we utilize timestamps whose log returns are outside the quantiles as price jump events.} 
        \label{subfig:log return}
     \end{subfigure}
     \hfill
     \begin{subfigure}[b]{0.5\textwidth}
        \centering
        \includegraphics[width=1\textwidth]{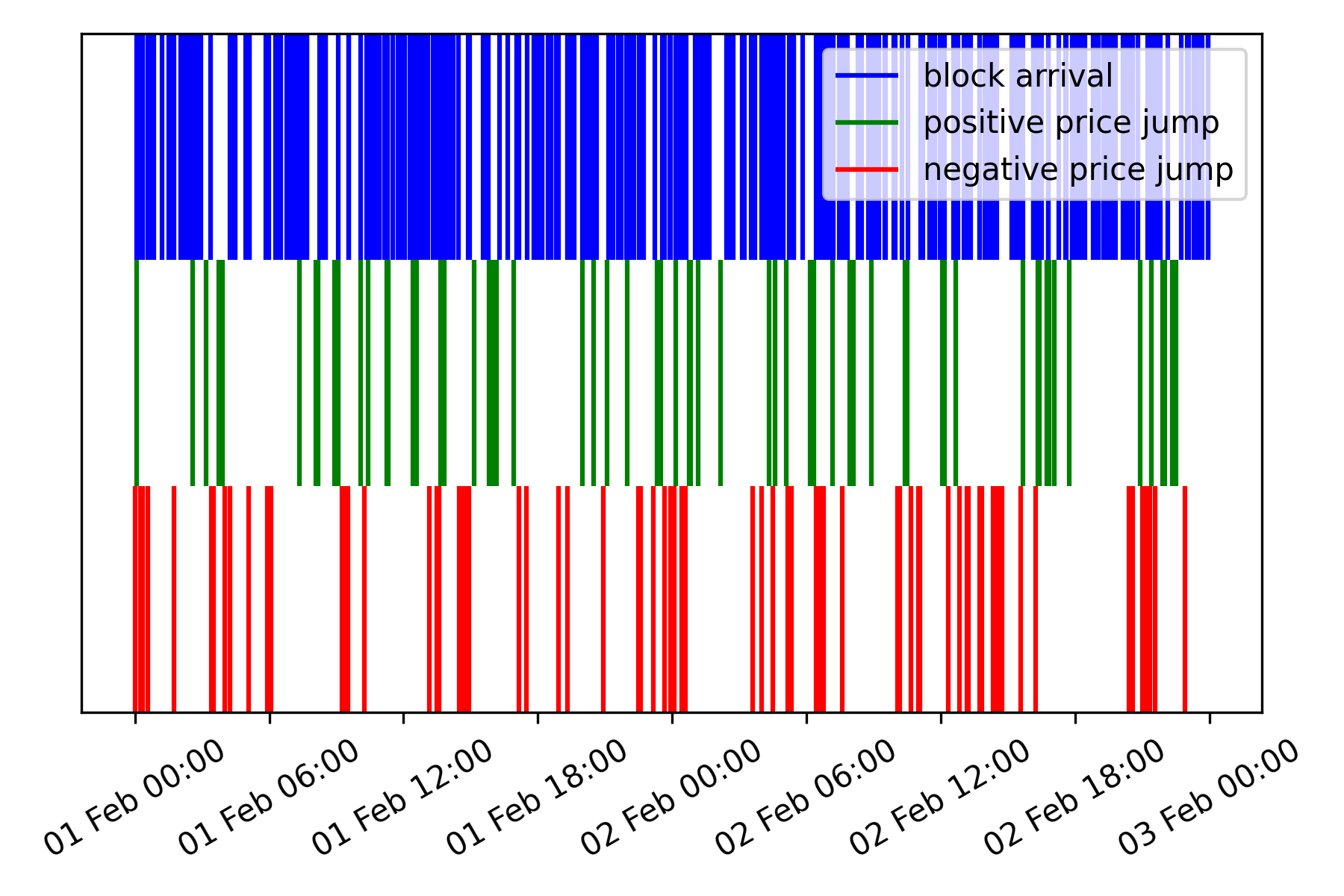}
        \caption{Illustration of the three types of events in the dataset -- block arrivals, positive bitcoin price jumps, and negative bitcoin price jumps -- between February 1 and February 3 of 2022.}
        \label{subfig:event}
    \end{subfigure}
    \label{fig:dataset}
    \caption{This figure contains two subfigures that show how the bitcoin price information was preprocessed to create positive and negative price jump events.}
\end{figure}

Due to the non-stationary behavior of bitcoin prices \cite{mudassir2020ML}, we consider the following time-varying thresholds based on rolling quantile. We look at a history of fixed length (3 hours) prior to the present data for each data point in the time series and compute the $10\%$ (resp. $90\%$) quantile of the history data as lower (resp. upper) threshold. If the current data is higher than the upper threshold, we include the time in the Hawkes process that represents positive price jumps. If it is lower than the lower threshold, we include the time in the Hawkes process representing negative price jumps.
Fig. \ref{subfig:log return} shows the price time series and the rolling quantiles of the log return. The log return that is outside the range of the quantiles will be regarded as a positive or negative price jump event. Fig. \ref{subfig:event} shows the three types of events -- block arrivals, positive price jumps, and negative price jumps -- during February 1 and February 3 of 2022. The price jump events transformed from price time series data, along with the block arrivals, are used to fit the Hawkes process model. 

\subsection{Model}
\label{subsec: block and crypto}

Let $N_1$ denote the counting process of block arrivals, $N_2$ (resp. $N_3$) denote the counting processes of positive (resp. negative) bitcoin price jumps. These three processes form a multivariate Hawkes process with conditional intensities of the form

\begin{equation}
	\lambda_i(t) = \mu_i + \sum_{j=1}^{3} \int_{0}^{t} \phi_{ij} (t-s) dN_j(s), \quad i=1,2,3
\end{equation}
where $\mu_i > 0$ denotes the background rate and $\phi_{ij}: (0, \infty) \rightarrow [0, \infty)$ denotes the kernel function used to model the cross-exciting effects, i.e., the change in the rate of occurrence of event $i$ caused by a previous realization of event $j$.  

As in (\ref{eq: intensity sumexp 1}, \ref{eq: intensity sumexp 2}), we utilize a sum of exponentials kernel with $U=3$ as the Hawkes process's kernel function
\begin{equation}
	\lambda_i(t) = \mu_i + \sum_{u=1}^{3} 
	\sum_{j=1}^{3} \int_0^t \alpha_{ij}^u e^{-\beta^u (t-s)} dN_j(s)
\end{equation}

The decay parameters $\{\beta^u\}_{u=1, 2, 3}$ are set to be $[2.340, 15.730, 21.875]$ by minimizing the log-likelihood using the Nelder-Mead algorithm\footnote{SciPy \url{https://docs.scipy.org/doc/scipy/index.html}.} with an initial value $[0.5, 5, 50]$. 
The parameters $[\alpha_{ij}^{u}]_{u=1, 2, 3}$ are obtained by maximum likelihood estimation\footnote{Tick\cite{bacry2017tick}: \url{https://x-datainitiative.github.io/tick/index.html}}

\begin{equation}
    [\alpha_{ij}^1] =
    \begin{bmatrix}  1.377 & 1.635 & 1.615\\  0.244 & 0.118 & 0.558\\  0.326 & 0.497 & 0.096
    \end{bmatrix}
\end{equation}
\begin{equation}
    [\alpha_{ij}^2] =
    \begin{bmatrix}  1.526 & 0. & 0.215\\  0. & 0. & 2.131\\  0. & 0.147 & 0.
    \end{bmatrix}
\end{equation}
\begin{equation}
    [\alpha_{ij}^3] =
    \begin{bmatrix}  0.02 & 0. & 0.\\  0. & 0. & 1.357\\  0. & 0. & 1.383
    \end{bmatrix}
\end{equation}

\vspace{0.2in}
\noindent
{\bf Interpreting the kernel norms of the fitted Hawkes process model:}
The kernel norms $||\phi_{ij}||_{\substack{i=1,2,3 \\ j=1,2,3}}$ are defined as
\begin{equation} \label{eq:kernel norm}
\begin{split}
    ||\phi_{ij}|| & = \int_{0}^{\infty} \phi_{ij}(t) dt \\
    & = \int_{0}^{\infty} \sum_{u=1}^{3} \alpha_{ij}^{u} e^{-\beta^u t} dt \\
    & = \sum_{u=1}^{3} \frac{\alpha_{ij}^{u}}{\beta^{u}}
\end{split}
\end{equation}

\begin{figure}
	\centering
	\includegraphics[width=0.5\textwidth]{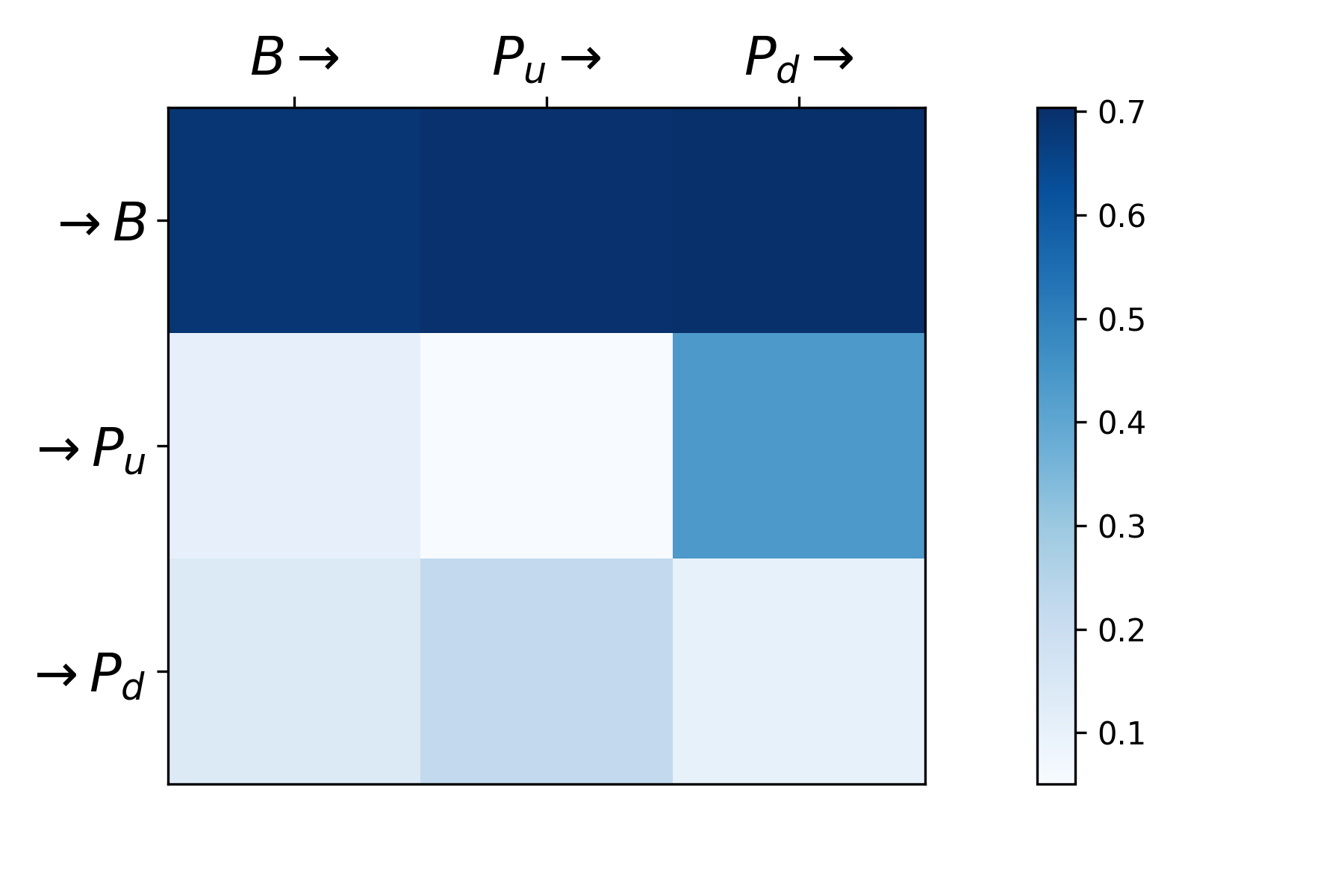}
	\caption{Kernel norms (\ref{eq:kernel norm}) of a Hawkes process fitted on the block arrivals and bitcoin price jumps data. $B$ counts the number of block arrivals and $P_u$ (resp. $P_d$) counts the number of upward (resp. downward) bitcoin price jumps. The main takeaway is that both positive and negative bitcoin price jumps boost the future block arrival rate, and bitcoin price returns demonstrate mean reversion.}
	\label{fig:Sum Exp Kernel Norms}
\end{figure}

The kernel norms $||\phi_{ij}||$ are the average number of events of type $i$ caused by an event of type $j$ \cite{bacry2016slowlyestimation}.
Fig. \ref{fig:Sum Exp Kernel Norms} shows the norms of the sum of exponentials kernel fitted to the dataset. 

% %%%%%%%%%%%%%%%%%%%%%%%%%%%%%%%%%%%%%%%%%%%%%%%%%%%%%%%%%%%%%%%%%%%%%%%%%%%%%%%%%%%%%%%%%%%
\vspace{0.2in}
\noindent
{\bf Goodness of Fit test:}
We test the model's goodness of fit to determine the appropriateness of the proposed Hawkes processes model for the crypto dataset. 
According to the random time change theorem\cite{daley2003introduction}, if $\{t_1, \cdots, t_k\}$ is a realisation over time $[0, T]$ from a point process with conditional intensity function $\lambda(\cdot)$, then the transformed points $\{\Lambda(t_1), \cdots, \Lambda(t_k) \}$ form a Poisson process with unit rate. Then the interarrival times $\{\Lambda(t_1), \Lambda(t_2) - \Lambda(t_1), \cdots, \Lambda(t_k) - \Lambda(t_{k-1}) \}$ should be independently and identically distributed as exponential random variable with mean $1$, i.e., $\textrm{exp}(1)$. Therefore, given a closed form expression of the compensator (\ref{eq:compensator multi sum}), we can compute the transformed points and use a quantile-quantile plot (Q-Q plot)  for exponential distribution to assess the quality of fit. 

\begin{figure}
     \centering
     \begin{subfigure}[b]{0.5\textwidth}
        \centering
        \includegraphics[width=1\textwidth]{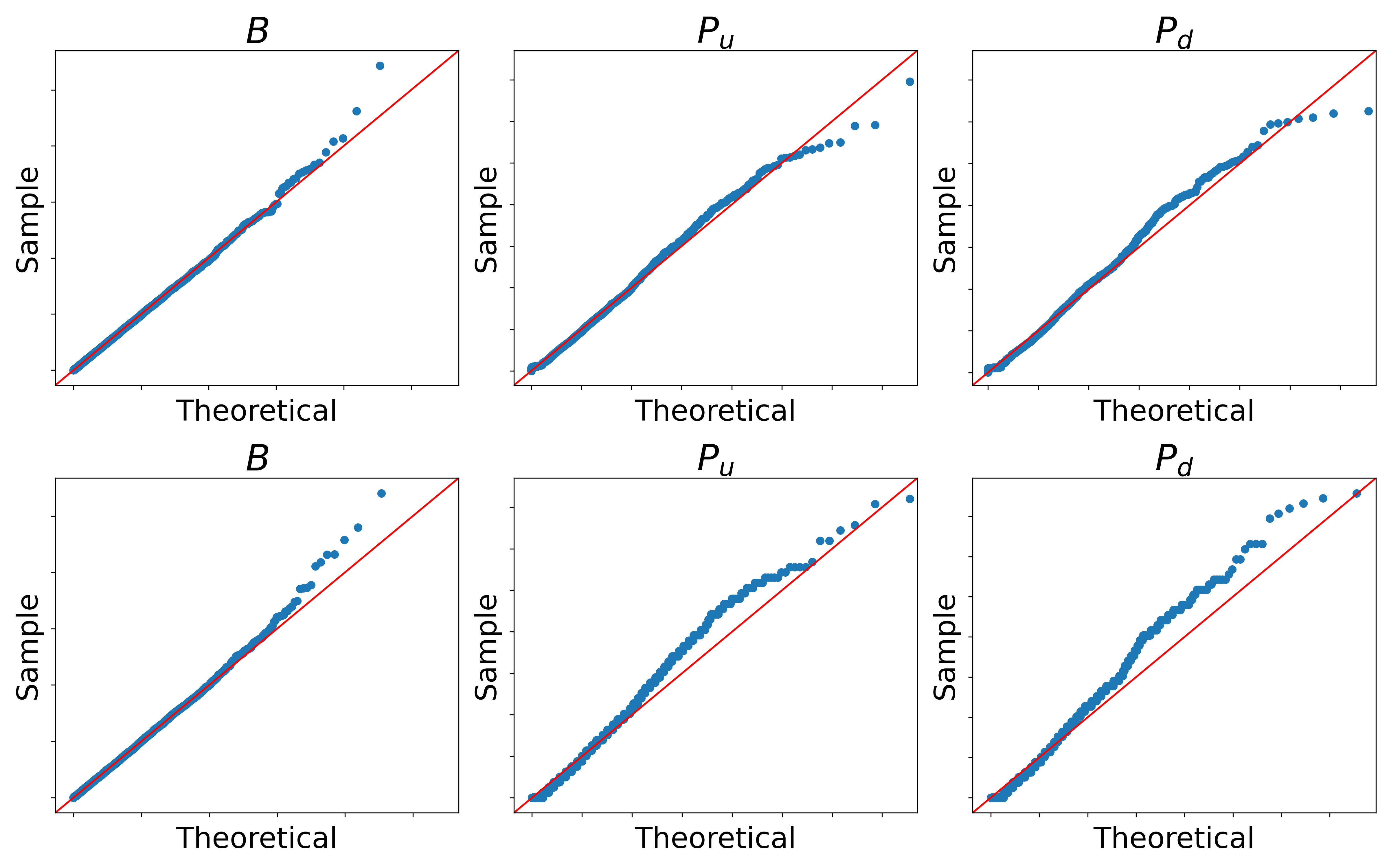}
        \caption{The subfigure shows the quantile-quantile (Q-Q) plots of the Hawkes process model (upper panel) and a homogeneous Poisson process model (lower panel). From left to right are the Q-Q plots of the block arrivals, positive price jumps, and negative price jumps. The timestamps are first converted to a unit-rate Poisson process using the random time change theorem, and the interarrival times are then compared to a $\textrm{Exp}(1)$ distribution.} 
        \label{subfig:QQ plot}
     \end{subfigure}
     \hfill
     \begin{subfigure}[b]{0.5\textwidth}
        \centering
        \includegraphics[width=1\textwidth]{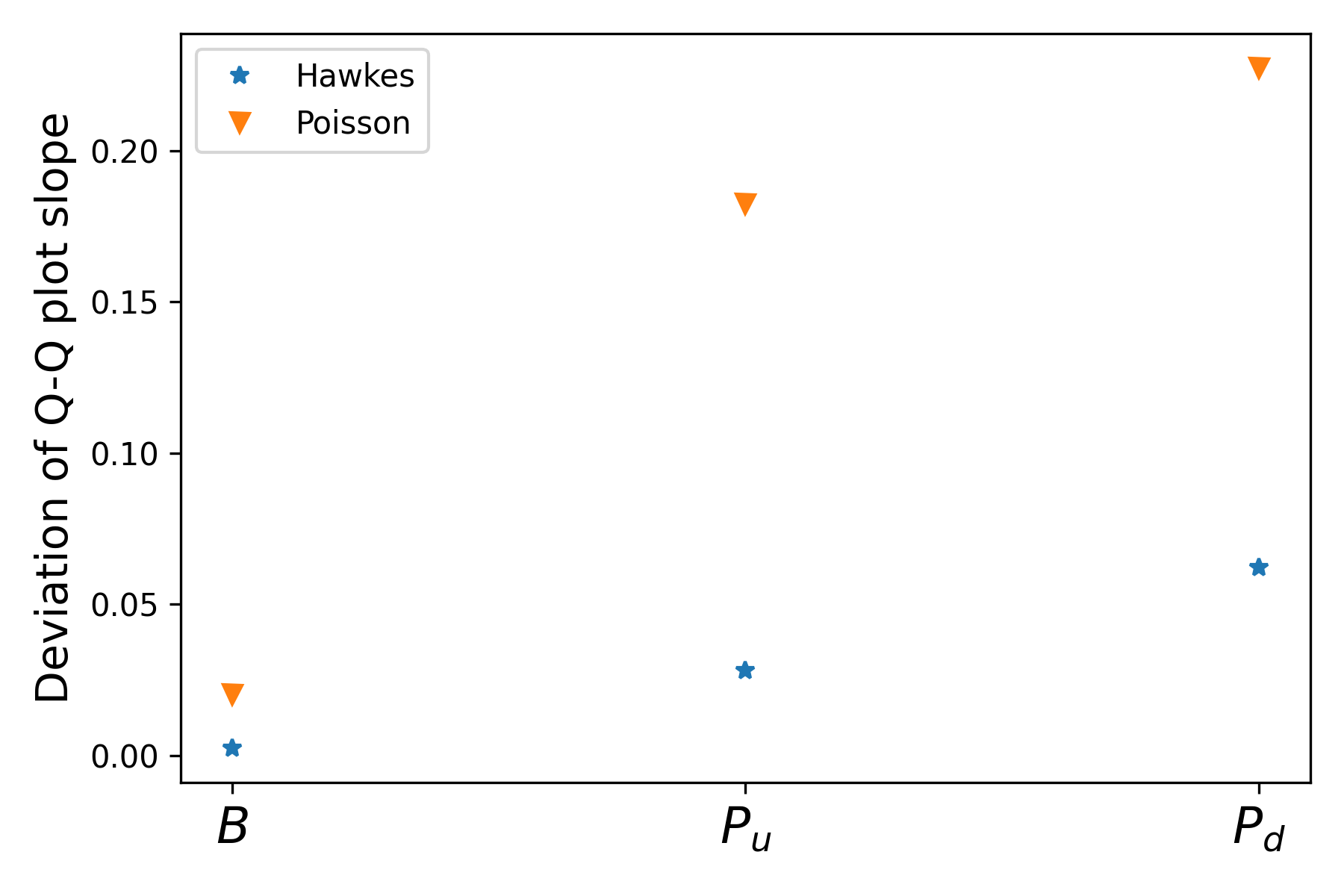}
        \caption{The subfigure shows the regression line's deviation from $45^{\circ}$ in the Q-Q plots of the Hawkes process and a Poisson process model. The smaller deviation of Hawkes process model indicates that it is a better fit to the bitcoin dataset.}
        \label{subfig:QQ deviation}
    \end{subfigure}
    \label{fig:QQ plot}
    \caption{The quantile-quantile (Q-Q) plots and slope deviations of the proposed Hawkes process model and a Poisson process model.}
\end{figure}

Fig. \ref{subfig:QQ plot} shows the Q-Q plot for the Hawkes process fit (upper panel) and the Q-Q plot for a homogeneous Poisson model (lower panel). Fig. \ref{subfig:QQ deviation} compares the deviation of the fitted line's slope from $45^{\circ}$. For modeling block arrivals, the Hawkes process model has a slope deviation of 0.002 whereas the Poisson process model has a slope deviation of 0.020; For modeling bitcoin price positive (resp. negative) jumps, the Hawkes process model has a slope deviation of $0.028$ (resp. 0.062) whereas the Poisson process model has a slope deviation of 0.182 (resp. 0.217).

\vspace{0.2in}
\noindent
{\bf Summary of Results: }
As seen in entries $(1, 2)$ and $(1, 3)$ of the kernel norms in  Fig. \ref{fig:Sum Exp Kernel Norms}, both positive and negative bitcoin price jumps increase the future block mining rate. The off-diagonal entries $(2, 3)$ and $(3, 2)$ represent mean reversion in bitcoin price returns: positive jumps will result in more negative jumps, while negative jumps will result in more positive jumps.

The smaller slope deviations as shown in Fig. \ref{subfig:QQ deviation} indicate that the Hawkes process model is a better fit to the bitcoin dataset than a Poisson process model. 

\section{Conclusions}
\label{sec:conclusion}
% \noindent
% {\bf Conclusions: }
% We proposed a blockchain enabled social media network which mitigates misinformation transmission. 
This paper investigated the use of multi-variate Hawkes process to model block arrivals in the bitcoin blockchain. Hawkes processes are self exciting point processes that can capture essential features of blockchain including the self- and cross-exciting impacts of block mining and bitcoin price volatility. The data from the Bitcoin price time series is converted into a sequence of positive and negative price jump events. Maximum likelihood estimation is then used to derive the kernel function for the trivariate Hawkes process. The kernel function norms indicate that price volatility encourages block mining, i.e., both positive and negative price jumps will boost block arrival rate. Using the random time change theorem, we show that the proposed Hawkes process model fits the bitcoin dataset better than a Poisson process model.

\appendices
\section{Block Data Cleaning} \label{append:cleaning}
All the results shown in Section \ref{sec:blockchain} are reproducible. The bitcoin price dataset can be downloaded from GitHub\footnote{\url{https://github.com/crypto-chassis/cryptochassis-data-api-docs}}, and the blockchain dataset can be obtained using Google Bigquery\footnote{\url{https://cloud.google.com/blog/topics/public-datasets/bitcoin-in-bigquery-blockchain-analytics-on-public-data}}. We will also make our code public when we submit the paper.

There are two forms of block timestamp errors in the query period, notably duplicated timestamps and out-of-order timestamps. In this section we explain how we clean these erroneous data.

\begin{figure}
	\centering
	\includegraphics[width=0.5\textwidth]{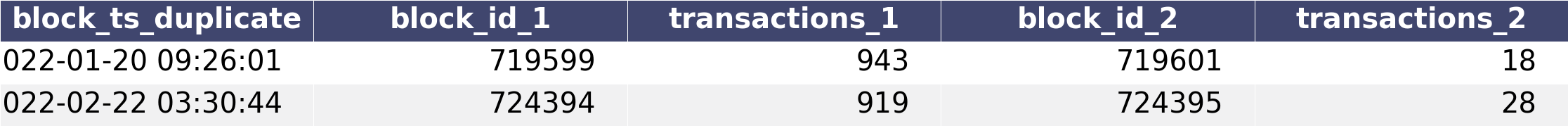}
	\caption{The table shows the blocks with duplicate time stamps in the query period. We preserve the blocks with higher number of transactions.}
	\label{fig:duplicate}
\end{figure}

\begin{figure}
	\centering
	\includegraphics[width=0.4\textwidth]{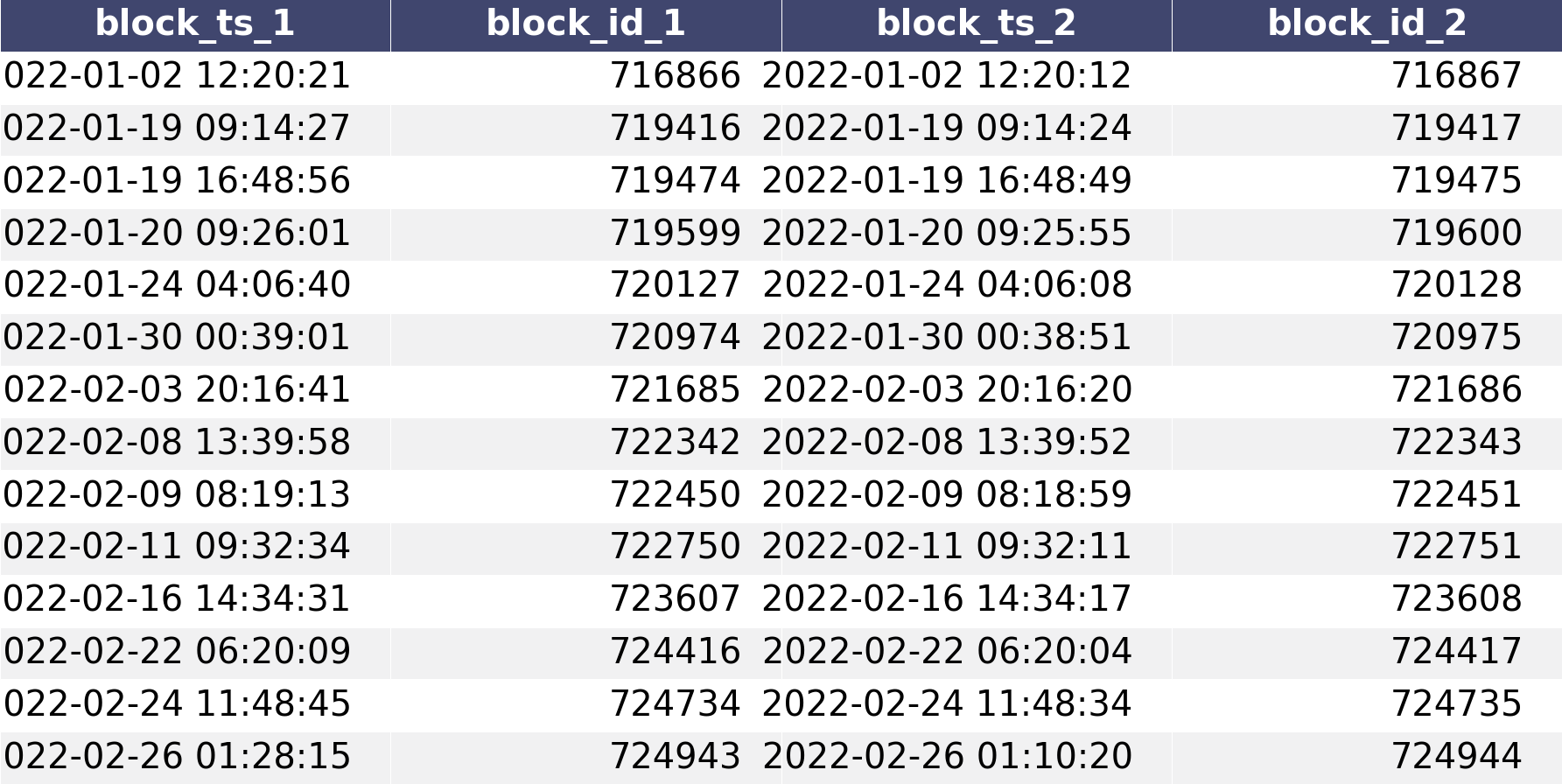}
	\caption{The table shows the blocks with out-of-order time stamps in the query period. We re-order the blocks by time stamps.}
	\label{fig:out of order}
\end{figure}

\begin{enumerate}
\item \textbf{Duplicated timestamps}: The first form of error is block timestamp duplication, which occurs when two blocks have the same timestamp. During the time period under consideration, there were two such instances (Fig. \ref{fig:duplicate}). One example is that block 719599 and block 719601 both arrived at 2022-01-20 09:26:01, we preserve block 719599 which contains more transactions and remove block 719601 from the dataset.

\item \textbf{Out-of-order timestamps}: The second form of error involves blocks whose timestamps are earlier than previous blocks. According to \cite{bowden2020modeling}, out-of-order timestamps are often caused by a miner using a timestamp from the future. In addition, some mining software and mining pools vary the timestamp to use it as an additional nonce in mining. During the query time period, there were 14 such instances (Fig. \ref{fig:out of order}).  We re-order the blocks by timestamps.

\end{enumerate}

% % use section* for acknowledgment
% \ifCLASSOPTIONcompsoc
%   % The Computer Society usually uses the plural form
%   \section*{Acknowledgments}
% \else
%   % regular IEEE prefers the singular form
% \fi

% This research was supported by the  U. S. Army Research Office under grants W911NF-19-1-0365, U.S. Air Force Office of Scientific Research under grant FA9550-22-1-0016, and the National Science Foundation under grant CCF-2112457.
% The authors would like to thank Yucheng Peng and Buddhika Nettasinghe for helpful discussions.

\ifCLASSOPTIONcaptionsoff
  \newpage
\fi



%\bibliographystyle{IEEEtran}

\bstctlcite{IEEEexample:BSTcontrol}
\bibliographystyle{IEEEtran}
\end{document}